\documentclass[aps,twocolumn,showpacs,superscriptaddress]{revtex4}
\usepackage{mathrsfs}
\usepackage{amssymb}
\usepackage{amsmath}
\usepackage{graphicx}
\usepackage{dcolumn}
\usepackage{bm}

\begin{document}

\title{The Euler Number of Bloch States Manifold and the Quantum Phases in
Gapped Fermionic Systems}
\author{Yu-Quan Ma}
\affiliation{School of Applied Science, Beijing Information Science and Technology
University, Beijing 100192, China}
\author{Shi-Jian Gu}
\affiliation{Department of Physics and ITP, The Chinese University of Hong Kong, Hong
Kong, China}
\author{Shu Chen}
\affiliation{Beijing National Laboratory for Condensed Matter Physics, Institute of
Physics, Chinese Academy of Sciences, Beijing 100190, China}
\author{Heng Fan}
\affiliation{Beijing National Laboratory for Condensed Matter Physics, Institute of
Physics, Chinese Academy of Sciences, Beijing 100190, China}
\author{Wu-Ming Liu}
\affiliation{Beijing National Laboratory for Condensed Matter Physics, Institute of
Physics, Chinese Academy of Sciences, Beijing 100190, China}
\date{\today}

\begin{abstract}
We propose a topological Euler number to characterize nontrivial topological
phases of gapped fermionic systems, which originates from the Gauss-Bonnet
theorem on the Riemannian structure of Bloch states established by the real
part of the quantum geometric tensor in momentum space. Meanwhile, the
imaginary part of the geometric tensor corresponds to the Berry curvature
which leads to the Chern number characterization. We discuss the topological
numbers induced by the geometric tensor analytically in a general two-band
model. As an example, we show that the zero-temperature phase diagram of a
transverse field XY spin chain can be distinguished by the Euler
characteristic number of the Bloch states manifold in a (1+1)-dimensional
Bloch momentum space.
\end{abstract}

\maketitle


\section{Introduction}

Since the discovery of the Berry phase \cite{Berry,Simon} contained in the
quantum state with cyclic adiabatic evolutions and the topological Chern
number interpretation for the adiabatic pumping \cite{Thouless,Niu} and the
quantized Hall conductance \cite{Laughlin,TKNN,Niu1984}, the geometric and
topological properties have been playing increasingly important roles in
quantum physics. Recently, new topological phases described by the $Z_{2}$
numbers have been found in quantum spin Hall effect and in topological
insulators \cite{Kane,Fu,Hasan,Qi}. A significant question is whether there
exists other quantum numbers to characterize the topological phases. In this
paper, we introduce a new topological quantum number--Euler number to
characterize the topological phase of a gapped fermionic ground state, which
is based on the Gauss-Bonnet theorem on the Riemannian structure established
by the real part of the quantum geometric tensor \cite{Provost,Berry1989} in
momentum space.

As a Hermitian metric induced on the quantum states manifold, the quantum
geometric tensor originates from defining a local $U(1)$ gauge invariant
quantum distance between two states in a parameterized Hilbert space. This
effort results in a Riemannian structure of the quantum states manifold, and
the corresponding Riemannian metric is given by the real part of the
geometric tensor. Remarkably, its imaginary part, canceled out in the
quantum distance, was later found to be just the Berry curvature up to a
constant coefficient.

The geometric tensor has recently drawn a lot of attention in characterizing
the novel collective behaviors of quantum many-body systems in low
temperature \cite{Resta,Haldane,Ma,Ryu,Rezakhani}. As a more general
covariant tensor than the Berry curvature on the Hilbert space geometry, the
quantum geometric tensor defined on the ground-state manifold is naturally
expected to shed some light on the understanding of quantum phase
transitions in many-body systems \cite{Sachdev}. Indeed, recent studies \cite%
{Zanardi,Venuti,Gu2010} have shown that the ground state geometric tensor
can provide a unified approach of the fidelity susceptibility \cite{Gu2007}
and the Berry curvature to demonstrate the singularity and scaling behaviors
exhibited in the vicinity of quantum critical point \cite%
{Pachos,Hamma,Zhu,Ma2009}.


On the other hand, the properties of the ground state, \emph{i.e., }some
physical response function, can be insensitive to local perturbations and
the system can undergo a topological phase transition \cite{Wen}, which is
beyond the Landau's second order phase transitions paradigm with the pattern
of symmetry breaking and local order parameters. One of the best known
examples is the integer quantum Hall effect, where the Hall conductivity is
expressed as the first Chern number in the units of $e^{2}/h$. To our
knowledge, the previous studies on the ground state geometric tensor are
trapped in the local properties, \emph{i.e., }the fidelity susceptibility
and the partial derivatives of Berry phase, and then only the phase
boundaries can be witnessed by this approach \cite%
{Abasto,Yang,Hamma2008,Garnerone}. Some works have show that the ground
state Berry phases protected by some symmetry can been used as local order
parameters to study various topological phases \cite{Hatsugai,Fufu}. The $Z_2
$ number characterization as a quantized Berry phase of Bloch states for
spin chain systems has been proposed in our recent work (see Ref.\cite%
{Ma2012}).

In this work, we introduce a topological Euler number, based on a quantum
geometric tensor defined in the Bloch states manifold, to characterize
quantum phases of a gapped fermionic ground state. We discuss this approach
analytically in a general two-band model. As an example, we show that there
exists a topological quantum phase transition in a transverse field XY
spin-1/2 chain in (1+1)-dimensional momentum space. We show that the phase
diagram can be distinguished by the topological Euler characteristic number,
meanwhile, a nontrivial $Z_{2}$ number is also obtained by the integral of
the Berry curvature as the imaginary part of the geometric tensor over half
of the Brillouin zone, which is converted from the first Chern number for
the time-reversal invariant Bloch Hamiltonian.

\section{Topological Euler number in momentum space}

To begin with, we introduce the notion of quantum geometric tensor in Bloch
momentum space, which can be derived from a gauge invariant distance between
two Bloch states on the $U(1)$ line bundle induced by the quantum adiabatic
evolution of the Bloch state $\left\vert {u}_{n}{\left( k\right) }%
\right\rangle $ of the $n$-th filled band. The quantum distance between two
states $\left\vert {u}_{n}{\left( k{+\delta k}\right) }\right\rangle $ {and }%
$\left\vert {u}_{n}{\left( k\right) }\right\rangle $ is given by $%
dS^{2}=\sum_{\mu ,\upsilon }\langle {{{\partial _{\mu }u}}}_{n}{{{\left(
k\right) dk^{\mu }}}\left\vert {{{\partial _{\upsilon }u}}}_{n}{{{\left(
k\right) dk^{\upsilon }}}}\right\rangle }$, where $\mu ,\nu $ denote the
components $k^{\mu }$ and $k^{\nu }$, respectively. The term $\left\vert {%
\partial _{\mu }u}_{n}\left( k\right) \right\rangle $ can be decomposed as $%
\left\vert {\partial _{\mu }u}_{n}{\left( k\right) }\right\rangle
=\left\vert {D_{\mu }u}_{n}{\left( k\right) }\right\rangle {+}\left[
\boldsymbol{1}-\mathcal{P}(k)\right] {{\left\vert {\partial _{\mu }u}_{n}{%
\left( k\right) }\right\rangle }}$, where $\mathcal{P}_{n}(k)=\left\vert {u}%
_{n}{\left( k\right) }\right\rangle \left\langle {u}_{n}{\left( k\right) }%
\right\vert $ is the projection operator and $\left\vert {D_{\mu }u\left(
k\right) }\right\rangle =\mathcal{P}_{n}(k)\left\vert {\partial _{\mu }u}_{n}%
{\left( k\right) }\right\rangle $ is the covariant derivative of $\left\vert
{u}_{n}{\left( k\right) }\right\rangle $ on the line bundle. Under the
condition of the quantum adiabatic evolution, the evolution of $\left\vert {u%
}_{n}{\left( k\right) }\right\rangle $ to $\left\vert {u}_{n}{\left(
k+\delta k\right) }\right\rangle $ will undergo a parallel transport, then
we have $\left\vert {D_{\mu }u}_{n}{\left( k\right) }\right\rangle =0$.
Finally, we can obtain $dS^{2}=\sum_{\mu ,\upsilon }\langle {\partial _{\mu
}u}_{n}{\left( k\right) }|\left[ \boldsymbol{1-}\mathcal{P}_{n}(k)\right]
\left\vert {\partial _{{\nu }}u}_{n}{\left( k\right) }\right\rangle {{%
dk^{\mu }dk^{\upsilon }}}${. }The quantum geometric tensor is given by%
\begin{equation}
Q_{\mu {\nu }}^{n}=\langle {\partial _{\mu }u}_{n}{\left( k\right) }|\left[
\boldsymbol{1-}\mathcal{P}_{n}(k)\right] \left\vert {\partial _{{\nu }}u}_{n}%
{\left( k\right) }\right\rangle .  \label{qgt}
\end{equation}%
The geometric tensor can be rewritten as $Q_{\mu {\nu }}^{n}=\mathcal{G}%
_{\mu {\nu }}^{n}-i\mathcal{F}_{\mu {\nu }}^{n}/2$, where $\mathcal{G}_{\mu {%
\nu }}^{n}=$Re$Q_{\mu {\nu }}^{n}\ $can be verified as a Riemannian metric,
or called the Fubini-Study metric, which establishes a Riemannian manifold
of the Bloch states, and then the quantum distance can be written as $%
dS^{2}=\sum_{\mu ,\upsilon }$Re$Q_{\mu {\nu }}^{n}{{dk^{\mu }dk^{\upsilon }}}
${.} The term{\ }$\mathcal{F}_{\mu {\nu }}^{n}=-2$\text{Im}$Q_{\mu {\nu }%
}^{n}$ is canceled out in the summation of the distance due to its
antisymmetry, but can associate to a $2$-form $\mathcal{F}^{n}=\sum_{\mu
,\upsilon }\mathcal{F}_{\mu {\nu }}^{n}{dk^{\mu }\wedge dk^{\nu }}$, which
is nothing but the Berry curvature. The geometric tensor is also a local
response function of the Bloch state, we find that the $\mathcal{G}_{\mu {%
\nu }}^{n}$ can be associated to a distance response as $\lim_{{\Delta k}%
\rightarrow 0}\left( {\Delta }S/{\Delta }k\right) ^{2}=\sum_{\mu ,\upsilon }%
\mathcal{G}_{\mu {\nu }}^{n}{\partial }_{k}k{^{\mu }}\otimes {\partial }%
_{k}k^{{\nu }}$ which is just the concept of fidelity susceptibility, and $%
\mathcal{F}_{\mu {\nu }}^{n}$ can be associated to a phase response function
as $\lim_{\sigma \rightarrow 0}\left( \left\vert {u}_{n}(k)\right\rangle _{%
\text{Final}}-\left\vert {u}_{n}(k)\right\rangle _{\text{Initial}}\right)
\left\vert {u}_{n}(k)\right\rangle _{\text{Initial}}^{-1}\sigma
^{-1}=\sum_{\mu ,\upsilon }\mathcal{F}_{\mu {\nu }}^{n}{\partial }_{k}{%
k^{\mu }\wedge \partial _{k}{k}}^{{\nu }}$, where $\sigma $ is the area
element enclosed by a cyclic path in the momentum space.

Now we consider the topological properties of the geometric tensor in
momentum space. Note that the Brillouin zone has the topology of torus if we
take the periodic gauge $\left\vert {u}_{n}{\left( k\right) }\right\rangle
=e^{iG\cdot r}\left\vert {u}_{n}{\left( k+G\right) }\right\rangle $, and
then the ground state satisfies $\left\vert {\Psi \left( k\right) }%
\right\rangle =\left\vert {\Psi \left( k+G\right) }\right\rangle $, where $G$
is the reciprocal lattice vector. Without loss of generality, we consider a
gapped fermionic Hamiltonian in 2D momentum space, where the unique Bloch
state $\left\vert {u}_{n}{\left( k\right) }\right\rangle $ forms a $U\left(
1\right) $ line bundle on a torus $T^{2}$ formed by the 2D Brillouin zone.
The corresponding Berry curvature $\mathcal{F}_{\mu {\nu }}^{n}$ for the
Bloch state $\left\vert {u}_{n}{\left( k\right) }\right\rangle $ is given by
$\mathcal{F}_{\mu {\nu }}^{n}=-2$\text{Im}$\langle {\partial _{\mu }u}_{n}{%
\left( k\right) }|\left[ \boldsymbol{1-}\left\vert {u}_{n}{\left( k\right) }%
\right\rangle \left\langle {u}_{n}{\left( k\right) }\right\vert \right]
\left\vert {\partial _{{\nu }}u}_{n}{\left( k\right) }\right\rangle $. The
topological invariant on the $U\left( 1\right) $ line bundle of all occupied
bands is the first Chern number
\begin{equation}
Ch_{1}=\frac{1}{2\pi }\sum_{n}\iint_{\text{Bz}}\mathcal{F}_{\mu {\nu }}^{n}d{%
k}^{\mu }d{k}^{{\nu }}.  \label{ch1}
\end{equation}%
What is more interesting is that there exists another topological invariant
---the Euler characteristic number, which originates from the Gauss-Bonnet
theorem on the 2D closed manifold established by the Riemannian metric $%
\mathcal{G}_{\mu {\nu }}^{n}$ for the Bloch state $\left\vert {u}_{n}{\left(
k\right) }\right\rangle $. The theorem states that the number $\chi \left(
\mathcal{M}^{2}\right) :=\frac{1}{2\pi }\int_{\mathcal{M}^{2}}\mathcal{K}dA$
is a topological invariant named Euler characteristic number and equals to $%
2\left( 1-g\right) $ with genus $g$ for a closed smooth manifold, where $%
\mathcal{K}$ is the Gauss curvature and $dA$ is the element of area of the
surface. In two dimensions, the Euler number of all occupied bands can be
calculated using the metric $\mathcal{G}_{\mu {\nu }}^{n}$ as follows
\begin{equation}
\chi =\frac{1}{4\pi }\sum_{n}\iint_{\text{Bz}}\mathcal{R}^{n}\sqrt{\det
\left( \mathcal{G}_{\mu {\nu }}^{n}\right) }d{k}^{\mu }d{k}^{{\nu }},
\label{eu}
\end{equation}%
where the $\mathcal{R}^{n}$ is the Ricci scalar curvature associate to the
Bloch state $\left\vert {u}_{n}{\left( k\right) }\right\rangle $ and here
the metric $\mathcal{G}_{\mu {\nu }}^{n}=$Re$\langle {\partial _{\mu }u}_{n}{%
\left( k\right) }|\left[ \boldsymbol{1-}\left\vert {u}_{n}{\left( k\right) }%
\right\rangle \left\langle {u}_{n}{\left( k\right) }\right\vert \right]
\left\vert {\partial _{{\nu }}u}_{n}{\left( k\right) }\right\rangle $. The
Ricci scalar curvature $\mathcal{R}$ can be calculated by the following
standard steps: $\mathcal{R}=\mathcal{G}^{ab}R_{ab}$, and $R_{ab}=R_{acb}^{c}
$, where the Riemannian curvature tensor $R_{abc}^{d}=\partial _{b}\Gamma
_{ac}^{d}-\partial _{a}\Gamma _{bc}^{d}+\Gamma _{ac}^{e}\Gamma
_{be}^{d}-\Gamma _{bc}^{e}\Gamma _{ae}^{d}$, and the Levi-Civit\`{a}
connection $\Gamma _{bc}^{a}$ can be calculated by $\Gamma _{bc}^{a}=\frac{1%
}{2}\mathcal{G}^{ad}\left( {\partial }_{b}\mathcal{G}_{dc}+{\partial }_{c}%
\mathcal{G}_{bd}-{\partial }_{d}\mathcal{G}_{cb}\right) $.

\section{Analytical results in two-band model}

Here let us consider a 1D translational invariant fermionic system with two
bands separated by a finite gap. The Hamiltonian can be written as $%
H=\sum_{l,l^{\prime }\in N}^{\text{PBC}}\Psi _{l,l^{\prime }}^{\dagger }%
\mathcal{H}_{l,l^{\prime }}\Psi _{l,l^{\prime }}$, where $\Psi _{l,l^{\prime
}}^{\dagger }=\left( c_{l}^{\dagger },c_{l^{\prime }}\right) $ denotes a
pair of fermionic creation and annihilation operators on the sites $l$ and $%
l^{\prime }$, $\mathcal{H}_{l,l^{\prime }}$ is a $2\times 2$ Hermitian
matrix, and the periodic boundary condition (PBC) has been imposed. In spite
of its simplicity, this model has a wide range of applications, such as the
Bogoliubov-de Gennes Hamiltonian in superconductivity and graphite systems.

In order to obtain an appropriate definition of geometric tensor to describe
the ground state of the system in a 2D manifold, we can perform the system a
local $U\left( 1\right) $ gauge transformation with $H(\varphi )=g(\varphi
)Hg(\varphi )^{+}$ by a time-depended twist operator $g(\varphi
)=\prod_{j=1}^{l-1}e^{i\varphi \left( t\right) c_{j}^{\dagger }c_{j}}$, and
here $\varphi \left( t\right) $ is a real function of time $t$. Note that
the terms $c_{l}^{\dagger }c_{l^{\prime }}^{\dagger }$ and $%
c_{l}c_{l^{\prime }}$ exist in the Hamiltonian $H$, so we have $\left[
g(\varphi ),H\right] \neq 0$ which ensure this operation is nontrivial.
Meanwhile, this operation does not change the system's energy spectrum, but
endows the system with the topology of a torus in (1+1)-dimensional momentum
space. After the Fourier transformations $c_{l}=\frac{1}{\sqrt{N}}%
\sum_{k}e^{ikl}a_{k}$ and $c_{l}^{\dagger }=\frac{1}{\sqrt{N}}%
\sum_{k}e^{-ikl}a_{k}^{\dagger }$, the Hamiltonian $H(\varphi )$ is
transformed into $H(\varphi )=\sum_{k\in \text{Bz}}\Psi _{k,\varphi
}^{\dagger }\mathcal{H}(k,\varphi )\Psi _{k,\varphi }$, where $\Psi
_{k,\varphi }=\left( a_{k,\varphi },a_{-k,\varphi }^{\dagger }\right) ^{T}$.
The Hamiltonian $\mathcal{H}(k,\varphi )$ can be generally written as $%
\mathcal{H}(k,\varphi )=\epsilon \left( k\right) $I$_{2\times
2}+\sum_{\alpha =1}^{3}d_{\alpha }\left( k,\varphi \right) \sigma ^{\alpha }$%
, where I$_{2\times 2}$ is the $2\times 2$ identity matrix and $\sigma
^{\alpha }$ are the three Pauli matrices, represent the pseudo-spin degree
of freedom. The energy spectrum is readily obtained as $E_{\pm
}(k)=\varepsilon \left( k\right) \pm \sqrt{\sum_{\alpha =1}^{3}d_{\alpha
}^{2}\left( k,\varphi \right) }$, and the corresponding eigenvector is%
\begin{equation}
u\left( \varphi ,k\right) _{\pm }=\frac{1}{\sqrt{2d\left( d\mp d_{3}\right) }%
}\left(
\begin{array}{c}
d_{1}-id_{2} \\
\pm d-d_{3}%
\end{array}%
\right) ,  \label{Blochfun}
\end{equation}%
where $d=\sqrt{\sum_{\alpha =1}^{3}d_{\alpha }^{2}\left( k,\varphi \right) }$%
. The Hamiltonian can be diagonalized as $H(\varphi )=\sum_{k\in \text{Bz}%
}E_{+}(k)\alpha _{k,\varphi }^{\dagger }\alpha _{k,\varphi }+E_{-}(k)\beta
_{k,\varphi }^{\dagger }\beta _{k,\varphi }$, where the quasi-particle
operators are $\alpha _{k,\varphi }=\left[ u\left( \varphi ,k\right) _{+}%
\right] ^{\dagger }\Psi _{k,\varphi }$ and $\beta _{k,\varphi }=\left[
u\left( \varphi ,k\right) _{-}\right] ^{\dagger }\Psi _{k,\varphi }$.

The ground state $\left\vert GS\right\rangle $ is the filled fermion sea $%
\left\vert GS\right\rangle =\prod_{k\in \text{Bz}}\beta _{-k,\varphi
}^{\dagger }\beta _{k,\varphi }^{\dagger }\left\vert 0\right\rangle $. Note
that if\ $\varphi \left( 0\right) =0$ and $\varphi \left( T\right) =\pi $,
then we have $H(0)=H(T)$ and if we adopt the periodic gauge $\left\vert {u}%
_{n}\left( k\right) \right\rangle =e^{iGl}\left\vert {u}_{n}\left(
k+G\right) \right\rangle $, $G$ is the reciprocal lattice vector. Then the
system has a topology of torus in the (1+1)-dimensional momentum space.
There exists a quantum geometric tensor $Q_{k\varphi }$ induced on the
momentum space $Q_{k\varphi }=\left\langle {\partial }_{k}u_{-}\right\vert %
\left[ 1-\left\vert u_{-}\right\rangle \left\langle u_{-}\right\vert \right]
\left\vert {\partial }_{\varphi }u_{-}\right\rangle $. More specifically,
substituting Eq. (\ref{Blochfun}) into the Berry curvature $\mathcal{F}%
_{k\varphi }=-2$\text{Im}$Q_{k\varphi }=\langle \partial _{k}u_{-}|\partial
_{\varphi }u_{-}\rangle -\langle \partial _{\varphi }u_{-}|\partial
_{k}u_{-}\rangle $, we can verify the relation that $\mathcal{F}_{k\varphi }=%
\frac{1}{2}\left[ \boldsymbol{\hat{d}}\cdot \partial _{k}\boldsymbol{\hat{d}}%
\times \partial _{\phi }\boldsymbol{\hat{d}}\right] _{k\varphi }$, $%
\boldsymbol{\hat{d}}$ denotes the unit vector $\boldsymbol{d/}d$. The
Riemannian metric $\mathcal{G}_{k\varphi }=$Re$Q_{k\varphi }=\frac{1}{2}%
\langle \partial _{k}u_{-}|\partial _{\varphi }u_{-}\rangle +\frac{1}{2}%
\langle \partial _{\varphi }u_{-}|\partial _{k}u_{-}\rangle $ $-\langle
\partial _{k}u_{-}|u_{-}\rangle \langle u_{-}|\partial _{\varphi
}u_{-}\rangle $. The calculation of $\mathcal{G}_{k\varphi }$ is tedious,
but we find that there exists a correspondence relation as $\sqrt{\det
\mathcal{G}}=\sqrt{\left[ \boldsymbol{\hat{d}}\cdot {\partial _{k}}%
\boldsymbol{\hat{d}}\times {\partial _{\varphi }}\boldsymbol{\hat{d}}\right]
^{2}}/4$, and then we have $\det \mathcal{G}=\left( \mathcal{F}/2\right)
^{2} $. Finally, we can calculate the Euler number as follows%
\begin{equation}
\chi =\frac{1}{16\pi }\int R\sqrt{\left( \boldsymbol{\hat{d}}\cdot {\partial
_{k}}\boldsymbol{\hat{d}}\times {\partial _{\phi }}\boldsymbol{\hat{d}}%
\right) ^{2}}dkd\varphi ,  \label{eu2}
\end{equation}%
and the first Chern number%
\begin{equation}
Ch_{1}=\frac{1}{4\pi }\int \left( \boldsymbol{\hat{d}}\cdot {\partial _{k}}%
\boldsymbol{\hat{d}}\times {\partial _{\phi }}\boldsymbol{\hat{d}}\right)
dkd\varphi .  \label{ch2}
\end{equation}

\section{XY spin chain in (1+1)-dimensional momentum space}

Here we choose an anisotropic XY spin-1/2 chain in a transverse magnetic
field as an example, the system is given by the following Hamiltonian $H_{%
\text{S}}=-\sum_{l\in N}^{\text{PBC}}\left( 1+\gamma \right)
s_{l}^{x}s_{l+1}^{x}+\left( 1-\gamma \right) s_{l}^{y}s_{l+1}^{y}+hs_{l}^{z}$%
, where $N\ $is the total sites of the spin chain, $\gamma $ is the
anisotropy parameter in the in-plane interaction and $h$ is the transverse
magnetic field. It is well known that the system undergoes a transition from
a paramagnetic to a ferromagnetic phase at $h=\pm 1$, which belongs to the
universality class of the transverse Ising model. The ferromagnetic order in
the XY plane is in the $x$-direction ($y$-direction) if $\gamma >0$ ( $%
\gamma <0$) and $|h|<1$. Now we subject the system to a local gauge
transformation $H_{\text{S}}(\varphi )=g(\varphi )H_{\text{S}}g(\varphi )^{+}
$ by a time-depended twist operator $g(\varphi )=\prod_{l}e^{i\varphi \left(
t\right) s_{l}^{z}}$, which in fact makes the system rotate on the spin
along the $z$-direction, so that we have $g(\varphi )s_{l}^{x}g(\varphi
)^{\dagger }=s_{l}^{x}\cos \varphi \left( t\right) -s_{l}^{y}\sin \varphi
\left( t\right) $ and $g(\varphi )s_{l}^{y}g(\varphi )^{\dagger
}=s_{l}^{y}\cos \varphi \left( t\right) +s_{l}^{x}\sin \varphi \left(
t\right) $. This operation extends the Hamiltonian into (1+1)-dimension
without changing its energy spectrum. Meanwhile, we assume\ $\varphi \left(
0\right) =0$, $\varphi \left( T\right) =\pi $ and we have $H_{\text{S}%
}(0)=H_{\text{S}}(T)$ and if we adopt the periodic gauge $\left\vert {u}%
_{n}\left( k\right) \right\rangle =e^{iGl}\left\vert {u}_{n}\left(
k+G\right) \right\rangle $, $G$ is the reciprocal lattice vector. Then the
system has a topology of torus $T^{2}$ in the (1+1)-dimensional Bloch
momentum space. After the standard calculation steps, we can transform the
spin Hamiltonian $H_{\text{S}}$ into a free fermion Hamiltonian as $\mathcal{%
H}\left( k,\varphi \right) =\epsilon \left( k,\varphi \right) $I$_{\text{2}%
\times \text{2}}+\sum_{\alpha =1}^{3}d_{\alpha }\left( k,\varphi \right)
\sigma ^{\alpha }$, where $\varepsilon \left( k,\varphi \right) =0$, $%
d_{1}\left( k,\varphi \right) =\frac{1}{2}\gamma \sin k\sin 2\varphi $, $%
d_{2}\left( k,\varphi \right) =\frac{1}{2}\gamma \sin k\cos 2\varphi $, and $%
d_{3}\left( k,\varphi \right) =-\frac{1}{2}\left( h+\cos k\right) $.
\begin{figure}[b]
\includegraphics[width=1.6in]{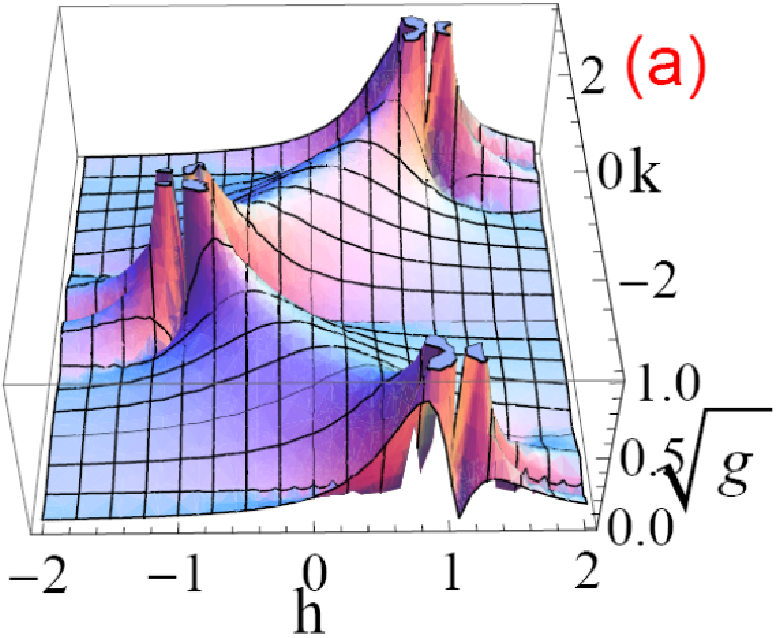} %
\includegraphics[width=1.6in]{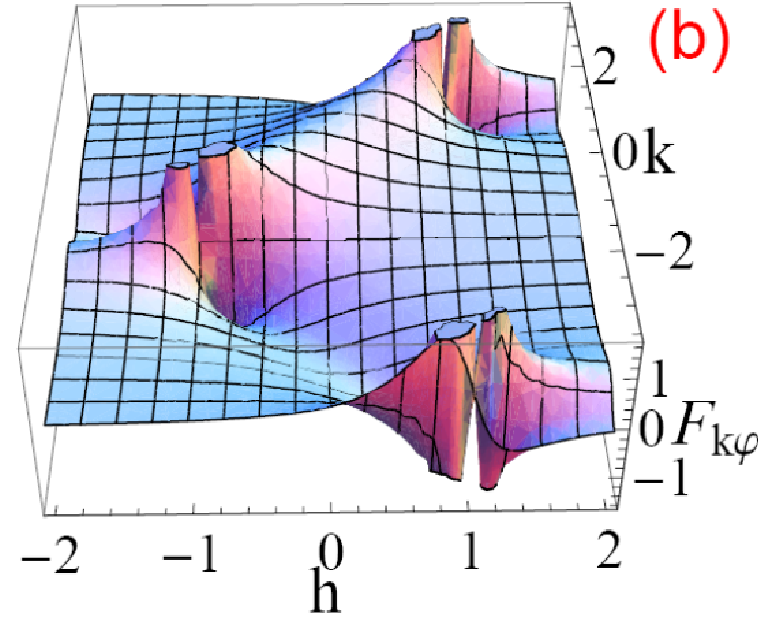}
\caption{(color online) (a) The gauge invariant $\protect\sqrt{g}$ as a
function of $k$ and $h$ ; (b) The Berry curvature $\mathcal{F}_{k\protect%
\varphi }$ as a function of $k$ and $h$.}
\label{local}
\end{figure}
\begin{figure}[t]
\begin{center}
\includegraphics[width=3in]{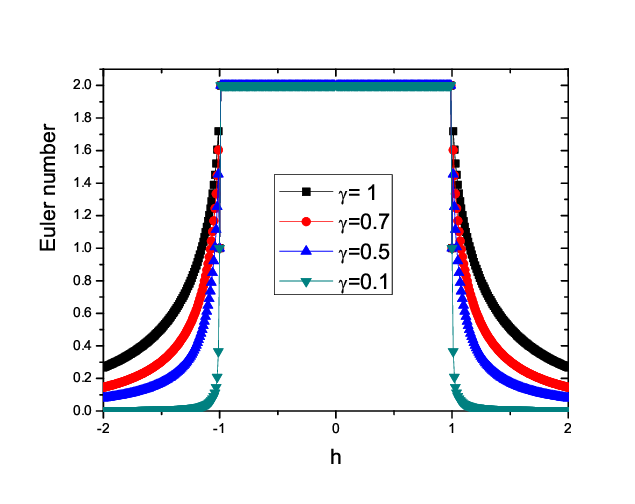}
\end{center}
\caption{(color online) The Euler number as a characterization of the phase
diagram of a transverse field XY spin chain as a function of the transverse
field $h$ and the anisotropy parameter $\protect\gamma $.}
\label{Euler}
\end{figure}
Here the Riemannian metric is given by
\begin{equation}
\mathcal{G}_{k\varphi }=\frac{1}{4}\left(
\begin{array}{cc}
\frac{\gamma ^{2}\left( 1+h\cos k\right) ^{2}}{\left( h+\cos k\right)
^{2}+\gamma ^{2}\sin ^{2}k} & 0 \\
0 & \frac{2\gamma ^{2}\sin ^{2}k}{\left( h+\cos k\right) ^{2}+\gamma
^{2}\sin ^{2}k}%
\end{array}%
\right) ,  \label{g}
\end{equation}%
and then we can obtain the Ricci scalar $R=8$ by the standard procedures.
Finally, the Euler number is given by
\begin{eqnarray}
\chi  &=&\frac{2}{\pi }\int \left\vert \frac{\boldsymbol{\vec{d}}\cdot {%
\partial _{k}}\boldsymbol{\vec{d}}\times {\partial _{\phi }}\boldsymbol{\vec{%
d}}}{d^{3}}\right\vert dkd\varphi   \notag \\
&=&\int_{0}^{\pi }\sqrt{\frac{\gamma ^{4}\left( 1+h\cos k\right) ^{2}\sin
^{2}k}{\left[ \left( h+\cos k\right) ^{2}+\gamma ^{2}\sin ^{2}k\right] ^{3}}}%
dk.
\end{eqnarray}

In Fig. (\ref{local}), we show the properties of the Berry curvature
and the metric tensor in the vicinity of quantum critical points. As
expected, both the Berry curvature and metric tensor exhibit
singularity around the phase transition points, but as a local
quantity, neither of them can serve as a topological order to
characterize a topological phase.

As shown in Fig. (\ref{Euler}), the ferromagnetic ordered phase in the XY
spin chain exhibits a nontrivial even Euler number $\chi =2$, and the Euler
number declines rapidly from $2$ to $0$ in the paramagnetic phase. Note that
the system is time reversal invariant, so the Berry curvature $\mathcal{F}%
_{k\varphi }$ is odd with $k$, and the first Chern number, as the integral
of the Berry curvature $\mathcal{F}_{k\varphi }$ in the Brillouin zone, is
equal to zero. It is clear that the Chern number is not a appropriate
topological number for this case. However, it has been pointed out in our
recent work \cite{Ma2012} that a non trivial $Z_{2}$ number can be obtained
as a quantized Berry phase along a loop over half (or quarter) of the
Brillouin zone,%
\begin{eqnarray}
Z_{2} &=&\frac{1}{2\pi }\int_{0}^{\pi }d\varphi \int_{0}^{\pi }\frac{\gamma
^{2}\left( 1+h\cos k\right) \sin k}{\left[ \left( h+\cos k\right)
^{2}+\gamma ^{2}\sin ^{2}k\right] ^{3/2}}dk  \notag \\
&=&\left\{
\begin{array}{lll}
1, &  & \text{if}\quad 0<\left\vert h\right\vert <1; \\
0, &  & \text{otherwise.}%
\end{array}%
\right. .
\end{eqnarray}%
Here, the $Z_{2}$ number is induced by the Berry curvature $\mathcal{F}%
_{k\varphi }$ ( imaginary part of the geometric tensor) reflecting the
topological obstruction of the $U(1)$ principal bundle. In contrast, the
Euler number is induced by the Riemannian metric $\mathcal{G}_{k\varphi }$ (
real part of the geometric tensor) of the Bloch state in a (1+1)-dimensional
momentum space, which reflects the number of the genus of the Bloch states
manifold.

\begin{figure}[tbh]
\begin{center}
\includegraphics[width=3in]{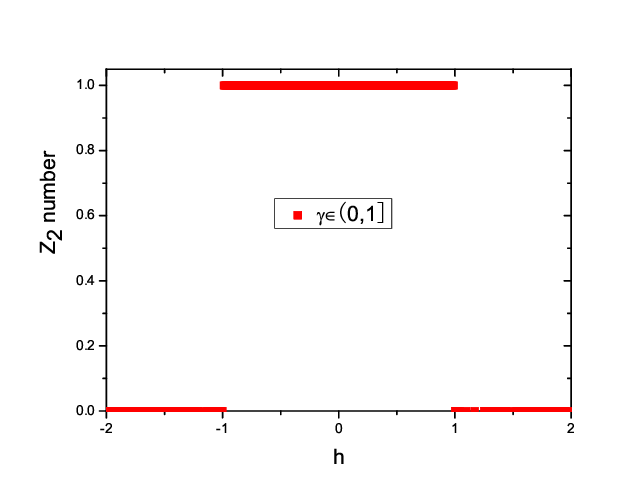}
\end{center}
\caption{(color online) (Left) The gapless point (red dot) enclosed in the
torus is converted into a monopole in the $\boldsymbol{d}$-space (Right),
and corresponds to a topological sphere of the ground state Riemannian
manifold with the Euler number $\protect\chi=2$.}
\label{illustration}
\end{figure}

The above results can be understood in an intuitional picture. As shown in
Fig. (\ref{illustration}), we can see the monopole as a gapless point in the
torus in the (1+1)-dimensional momentum space corresponding to the zero
point in the three dimensional $\boldsymbol{d}$-space. It can be verified
that only if $\left\vert h\right\vert <1$ then the monopole is enclosed by
the surface of $d_{\alpha }\left( k,\varphi \right) $, which corresponds to
a topological sphere of the ground state Riemannian manifold with the Euler
number $\chi=2$. In this case, the first Chern number is not an effective
characterization because of the time reversal invariance, and the $Z_2$
number can be expressed as the quantized upper hemispherical flux of the
monopole.

\section{Conclusion}

We have introduced the Euler characteristic number $\chi $ as a new
topological quantum number to distinguish the topological phase transitions
in gapped fermionic systems. In particular, we show that the
zero-temperature phase diagram of a transverse field XY spin chain can be
characterized by the Euler numbers in the (1+1)-dimensional momentum space.
We show that it is the Euler number instead of the Chern number to be an
effective characterization of the nontrivial topological phases in the
time-reversal invariant systems. This approach provides another description
of the topological nature of the ground state in gapped fermionic systems.
We hope that this work will raise renewed interest in the understanding of
the topological nature in quantum condensed-matter systems.

\emph{Note added.}---\ After this manuscript has been submitted, we note
that recently a similarly work on the metric tensor and the Euler
characteristic number appeared \cite{Kolodrubetz}.

\section{Acknowledgments}

This work is supported by the NKBRSFC under grants Nos. 2011CB921502,
2012CB821305, 2010CB922904, 2009CB930701, NSFC under grants Nos. 10934010,
60978019, NSFC-RGC under grants Nos. 11061160490, 1386-N-HKU748/10, and the
RGC of HKSAR, China (Project No. HKUST3/CRF/09).

\end{document}